\documentclass[reprint,superscriptaddress,amsmath,amssymb,aps,prl]{revtex4-2}
\usepackage{graphicx}
\usepackage{dcolumn}
\usepackage{bm}
\usepackage{color}
\usepackage{units}
\usepackage{wasysym}
\usepackage{MnSymbol}
\usepackage{comment}
\usepackage{times}
\usepackage[unicode=true,pdfusetitle,bookmarks=false,colorlinks=true,citecolor=blue,urlcolor=blue,linkcolor=blue]{hyperref}
\begin{document}

\title{Phase Diagram of Active Brownian Spheres: Crystallization and the Metastability of Motility-Induced Phase Separation}

\author{Ahmad K. Omar}
\email{aomar@berkeley.edu}
\affiliation{Department of Materials Science and Engineering, University of California, Berkeley, California 94720, USA}
\affiliation{Department of Chemistry, University of California, Berkeley, California 94720, USA}
\author{Katherine Klymko}
\affiliation{Computational Research Division, Lawrence Berkeley National Laboratory, Berkeley, California 94720, USA}
\affiliation{NERSC, Lawrence Berkeley National Laboratory, Berkeley, California 94720, USA}
\author{Trevor GrandPre}
\affiliation{Department of Physics, University of California, Berkeley, California 94720, USA}
\author{Phillip L. Geissler}
\email{geissler@berkeley.edu}
\affiliation{Department of Chemistry, University of California, Berkeley, California 94720, USA}
\affiliation{Chemical Sciences Division, Lawrence Berkeley National Laboratory, Berkeley, California 94720, USA}

\begin{abstract}
Motility-induced phase separation (MIPS), the phenomenon in which purely repulsive active particles undergo a liquid-gas phase separation, is among the simplest and most widely studied examples of a nonequilibrium phase transition. Here, we show that states of MIPS coexistence are in fact only metastable for three-dimensional active Brownian particles over a very broad range of conditions, decaying at long times through an ordering transition we call active crystallization. At an activity just above the MIPS critical point, the liquid-gas binodal is superseded by the crystal-fluid coexistence curve, with solid, liquid, and gas all coexisting at the triple point where the two curves intersect. Nucleating an active crystal from a disordered fluid, however, requires a rare fluctuation that exhibits the nearly close-packed density of the solid phase. The corresponding barrier to crystallization is surmountable on a feasible timescale only at high activity, and only at fluid densities near maximal packing. The glassiness expected for such dense liquids at equilibrium is strongly mitigated by active forces, so that the lifetime of liquid-gas coexistence declines steadily with increasing activity, manifesting in simulations as a facile spontaneous crystallization at extremely high activity.  
\end{abstract}

\maketitle

\textit{Introduction.--} The equilibrium crystallization of hard spheres~\cite{Alder1957} is the canonical example of entropically driven ordering of particle configurations:
For a range of volume fractions $\phi$, a fluid of hard spheres in three dimensions (3D) undergoes a symmetry breaking transition into coexisting disordered (fluid) and ordered (solid) phases ~\cite{Hoover1968, Pusey1986, Pusey1989, Rintoul1996, Auer2001, Torquato2002, Zaccarelli2009, Pusey2009, Richard2018a, Richard2018}.
Boltzmann statistics provide an unambiguous physical interpretation of the driving force for this transition: the free volume generated by ordering permits a more diverse set of particle configurations, whose entropy is the sole contribution to the free energy of hard spheres.
This order-disorder transition is entirely geometric in origin and is controlled solely by $\phi$. 

The influence of \textit{nonconservative} dynamics on the melting transition of hard spheres is an open and important subject in nonequilibrium statistical mechanics: How do driven dynamics compete with entropic geometric forces to create or destroy order?
To this end, active Brownian particles (ABPs) have emerged as a paradigmatic minimal model of driven systems and have aided in advancing our general understanding of nonequilibrium phase behavior~\cite{CatesMichaelE.AndTailleur2015, Bechinger2016, Mallory2018}.
In the athermal limit, the ABP model has only two distinct control parameters, both geometric in character. Dimensional analysis reveals one as $\phi$ and the other as the ratio of the persistence (or ``run") length of a free particle's trajectory $\ell_0$ to the particle size $\sigma$.   
This run length provides a convenient and direct measure of the time-irreversible motion of active particles, allowing for a continuous departure from reversible dynamics ($\ell_0 /\sigma \rightarrow 0$)~\cite{Fodor2016, Speck2016, Mandal2017} where equilibrium hard-sphere physics should be precisely recovered.

Further motivating the study of active crystallization is the knowledge that for finite run lengths, ABPs exhibit a distinct geometric transition that has garnered considerable interest: the so-called motility-induced phase separation (MIPS)~\cite{FilyYaouenandMarchetti2012, Redner2013, Buttinoni2013, CatesMichaelE.AndTailleur2015}. 
This uniquely nonequilibrium phenomenon requires no interparticle attraction, yet appears to be a genuine liquid-gas transition, with no evidence of rotational symmetry breaking between the coexisting phases in 3D~\cite{Wysocki2014, Stenhammar2014, Nie2020, Omar2020a, Turci2021}. 
The apparent and conspicuous absence of an ordered phase for activities in the vicinity of the MIPS phase boundary raises the intriguing question: Does the crystallization transition disappear as the system departs from equilibrium?

In this Letter we aim to clarify the relationship between MIPS and crystallization out of equilibrium.
To this end, we present results of extensive simulations of active Brownian hard spheres, conducted over a broad range of conditions.
The majority of computational work on ABP ordering transitions has focused on repulsive disks in 2D~\cite{Bialke2012, Fily2014, Cugliandolo2017, Digregorio2018, Klamser2018, Caporusso2020, Caprini2020b}, where the relationship between MIPS and crystallization is obscured by complications that long muddied the nature of freezing even for hard disks at equilibrium~\cite{Mermin1966, Bernard2011}.
We instead construct phase diagrams for active Brownian hard spheres in 3D, where the order-disorder transition is straightforward in the equilibrium limit.
These results reveal that the crystallization coexistence region in fact \textit{expands} with increasing activity, engulfing the MIPS phase boundary everywhere except for a narrow range of control parameters. 
Slightly above the critical activity, the solid-fluid phase boundary intersects the liquid-gas binodal, forming an active triple point where solid, liquid, and gas may coexist.
The proximity of the triple and critical points renders nearly the entirety of the MIPS phase boundary \textit{metastable}, with solid-fluid coexistence being the globally stable configuration. 
The frequent observation of liquid-gas coexistence (and its apparent stability) is due to the remarkably narrow region of the phase diagram where nucleation of an active crystal from a disordered fluid can be readily observed.
By locating these regions, we are able to identify the rate-limiting features of the active crystal nucleation landscape.

\textit{Model system.--} We consider the simplest active system that captures the equilibrium crystallization limit for vanishing activity: 3D active Brownian hard spheres. 
Each of the $N$ particles experiences three forces: a drag force $-\zeta \mathbf{\dot{x}}$ proportional to the particle velocity $\mathbf{\dot{x}}$, a conservative (pairwise) interparticle force $\mathbf{F^C}[\mathbf{x}^N]$, where $\mathbf{x}^N$ is the set of all particle positions, and an active self-propelling force $\mathbf{F^{A}}=\zeta U_0 \mathbf{q}$. 
The particle orientations $\mathbf{q}$ independently obey diffusive 3D rotary dynamics $\mathbf{\dot{q}} = \mathbf{\Omega} \times \mathbf{q}$ where the stochastic angular velocity has mean $\mathbf{0}$ and variance $\langle \mathbf{\Omega}(t) \mathbf{\Omega}(0) \rangle = 2/\tau_R \delta(t) \mathbf{I}$ and $\tau_R$ is the characteristic reorientation time.
We take the interparticle force $\mathbf{F^C}[\mathbf{x}^N; \varepsilon, \sigma]$ to result from a Weeks-Chandler-Anderson potential~\cite{Weeks1971} (characterized by a Lennard-Jones diameter $\sigma$ and energy $\varepsilon$) and take $\zeta U_0$, $\sigma$, and $\tau_R$ to be the characteristic units of force, length and time, respectively. 
The overdamped Langevin equation for the dimensionless velocity $\overline{\mathbf{\dot{x}}}$ naturally follows as
\begin{equation}
\label{eq:eom}
\overline{\mathbf{\dot{x}}} = \frac{\ell_0}{\sigma} \left ( \mathbf{q} + \mathbf{\overline{F}^{C}}[\mathbf{\overline{x}}^N; \mathcal{S}] \right ), 
\end{equation}
where $\ell_0 = U_0 \tau_R$. The dimensionless force $\mathbf{\overline{F}^{C}}$ depends on the reduced positions $\mathbf{\overline{x}}^N$ and is fully characterized by the ``stiffness" parameter $\mathcal{S} \equiv \varepsilon/(\zeta U_0\sigma)$.

Despite our use of a continuous potential, the hard-sphere limit is very closely approached in these simulations.
Lacking translational Brownian motion (which attenuates the influence of activity on the phase behavior~\footnote{Thermal energy $\protect k_{\protect B}T$ adds an
  additional dimension to the phase diagram that measures its strength relative
  to activity, $\protect \mathcal {T}=k_{\protect B}T /(\zeta U_0\sigma )$.
  In the limit that $\protect \mathcal {T}\rightarrow \infty $, thermal effects
  dominate, and active phase behavior vanishes for all $\ell _0/\sigma $. Our
  study aims to establish the active (i.e.,~athermal $\protect \mathcal {T}=0$)
  limit.}), and inertia (which also profoundly alters active phase behavior~\cite{Mandal2019}), these particles strictly exclude volume with a diameter $d$ set by $\mathcal{S}$.
Continuous repulsions act only at distances between $d$ and $2^{1/6} \sigma$, a range that quickly becomes negligible as $\mathcal{S}$ increases.
We use a stiffness $\mathcal{S}=50$ for which $d/(2^{1/6}\sigma) = 0.9997$, effectively achieving hard-sphere statistics.
Holding $\mathcal{S}$ to remain in this hard-sphere limit, the system state is independent of the active force magnitude and is fully described by two geometric parameters: the volume fraction $\phi = N\pi (2^{1/6} \sigma)^3/6V$ and the dimensionless intrinsic run length $\ell_0 / \sigma$.

All simulations were conducted with a minimum of $54000$ particles using \texttt{HOOMD-blue}~\cite{Anderson2020, Note2}.

\begin{figure}
	\centering
	\includegraphics[width=.475\textwidth]{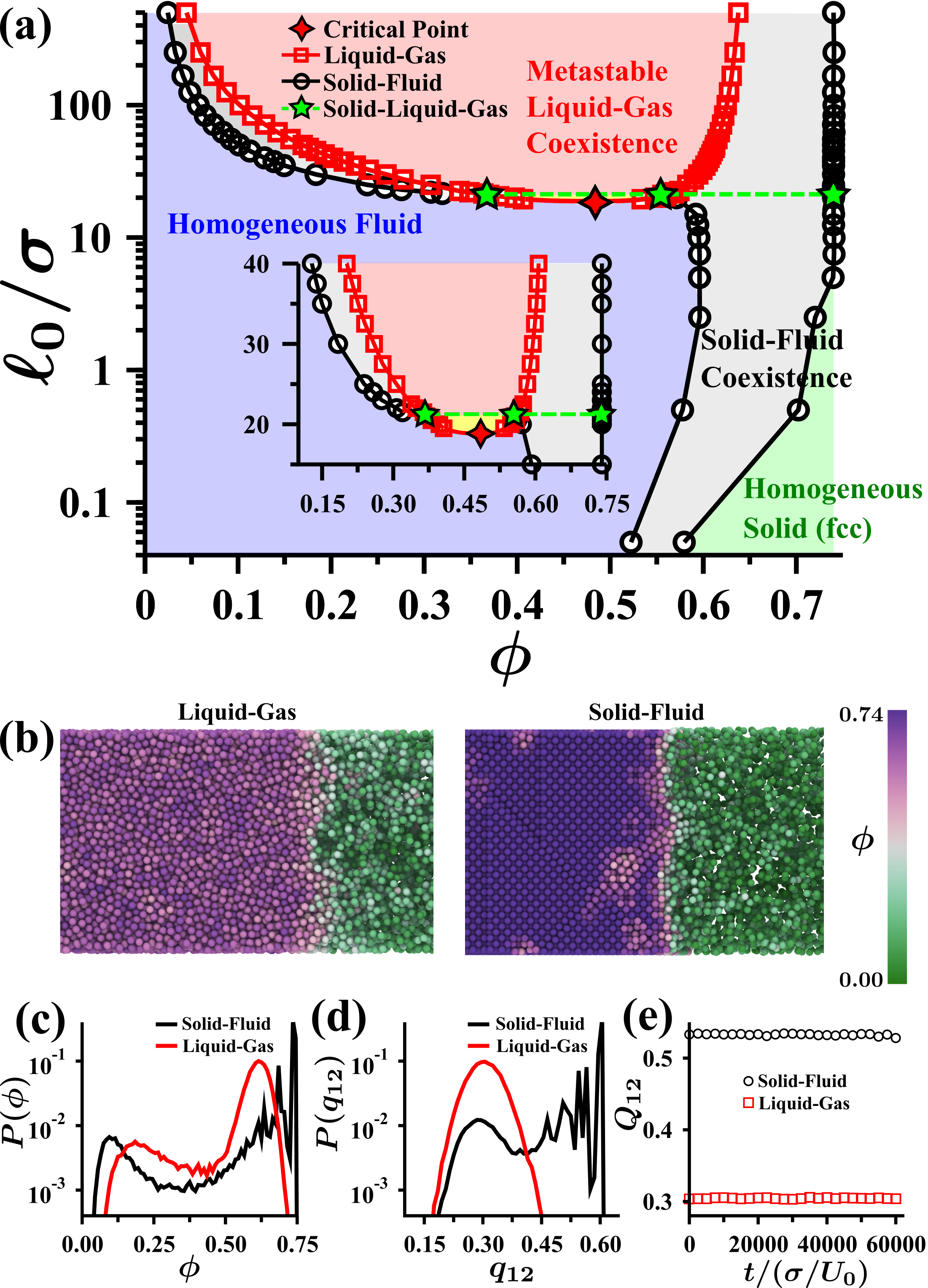}
	\caption{\protect\small{{(a) Phase diagram of 3D active hard spheres, with the critical region magnified in the inset. For $(\ell_0 / \sigma = 50, \phi = 0.5)$, (b) representative configurations of liquid-gas and solid-fluid coexistence. Corresponding probability distributions for (c) local volume fraction (using particle Voronoi volumes), and (d) $q_{12}$ (which takes a value of $q_{12}\approx 0.6$ for perfect fcc order and $q_{12}\approx 0.3$ for a disordered fluid). (e) Global symmetry parameter $Q_{12}$ as a function of time for both coexistence scenarios.}}}
	\label{figure1}
\end{figure}

\begin{figure*}
	\centering
	\includegraphics[width=.95\textwidth]{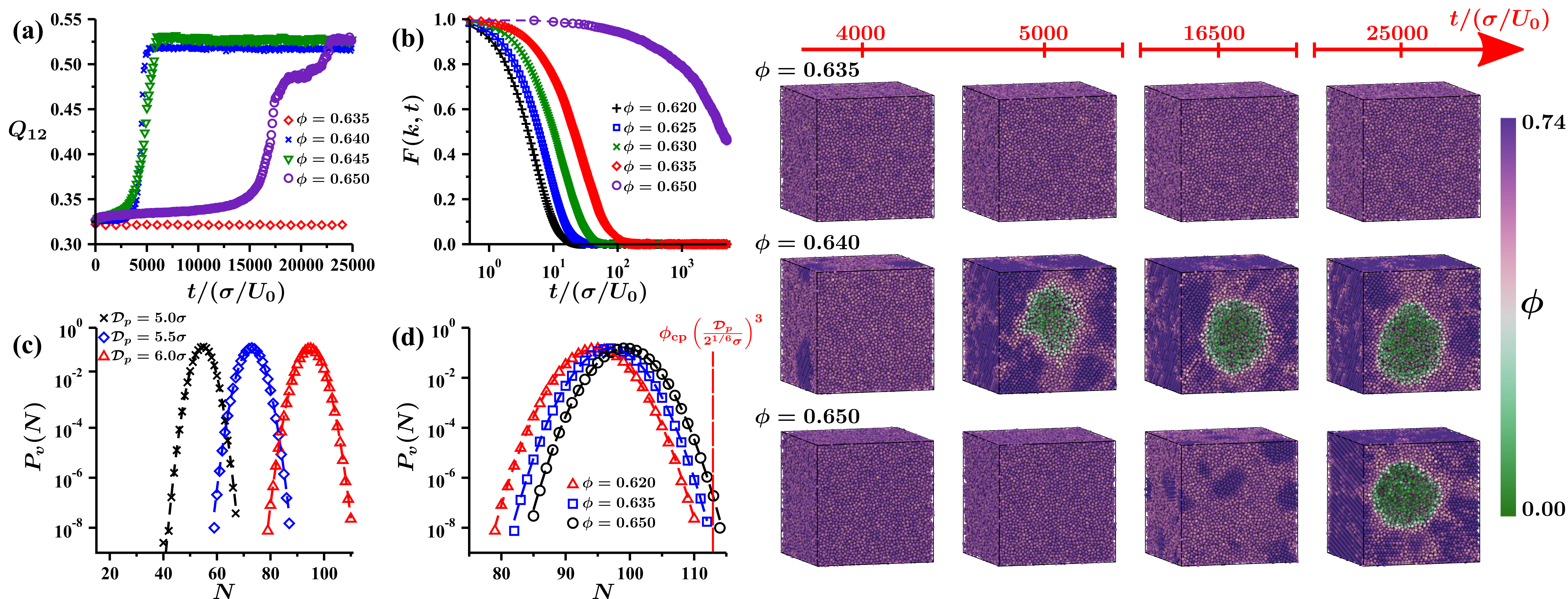}
	\caption{\protect\small{{Crystal nucleation from metastable active fluids with $\ell_0 / \sigma = 50$. (a) Time evolution of $Q_{12}$ and accompanying snapshots (right). (b) Dynamic structure factor $F(k,t)$ evaluated at the wavevector $k=|\mathbf{k}| = k^*$ corresponding to the first peak of $F(k,0)$~\cite{Note2}. Probe volume occupation probability $P_v(N)$ plotted as a function of $N$ in (c) for $\phi = 0.62$ at various probe diameters $\mathcal{D}_p$, and in (d) for $\mathcal{D}_p = 6.0\sigma$ at various densities. Lines are Gaussian distributions with the same mean and variance as simulation data.}}}
	\label{figure2}
\end{figure*}

\textit{Phase diagram.--} The phase diagram of 3D active hard spheres is presented in Fig.~\ref{figure1}.
Initially homogeneous~\cite{Note2} systems prepared within the liquid-gas binodal are often observed to spontaneously phase separate, the widely reported MIPS.
For all activities within the 3D MIPS phase boundary, the coexisting phases differ only in density, appearing to 
share precisely the same symmetry, shown in Fig.~\ref{figure1}(b).
More quantitatively, Figs.~\ref{figure1}(c) and (d) show the probability distribution of local density to be bimodal, while  $q_{12}$ (the per-particle Steinhardt-Nelson-Ronchetti order parameter~\cite{Steinhardt1983} measuring 12-fold rotational symmetry) is Gaussian distributed to a good approximation.

The critical point associated with this liquid-gas transition is found by assuming critical scaling of the order parameter, which we take to be the difference between liquid- and gas-phase densities $\phi_{\rm liq}-\phi_{\rm gas}$. 
Defining the reduced activity as $\tau = \frac{\ell_0-\ell_c}{\ell_c}$, the order parameter is anticipated to scale as $\phi_{\rm liq}-\phi_{\rm gas} \propto \tau^{\beta}$ ($\tau > 0$).
By fitting the coexisting densities nearest to the critical point~\cite{Note2}, we extract a critical activity $\ell_c/\sigma \approx 18.8$ and critical exponent $\beta \approx 0.33$. The latter value agrees suggestively (and perhaps fortuitously) with the 3D Ising universality class. 
A full critical scaling analysis~\cite{Binder1981, Binder1987, Rovere1988, Rovere1990, Rovere1993} (such as those recently performed on 2D active systems~\cite{Siebert2018, Partridge2019, Maggi2021, Dittrich2020, Adachi2020}) will be required to confirm the robustness of this apparent agreement. 
The critical density is found to be $\phi_c \approx 0.483$.

The order-disorder transition, by contrast, is notably absent from the literature on 3D ABPs, a direct consequence of formidable nucleation barriers that will be described below (see Fig.~\ref{figure2}).
To access this transition, we devise a simulation protocol~\cite{Note2} that biases the system to form face-centered-cubic (fcc) crystals, later established as the stable ordered phase for this system.
In a nutshell, we initialize the particles in a perfect fcc configuration at $\phi = 0.7$ and perform a uniaxial extension to sweep through $\phi$ and identify regions of solid-fluid coexistence. 
Long simulations~\cite{Note2} are run to verify the stability of the observed coexistence.
The resulting coexisting solid and fluid densities are reported in our phase diagram [Fig.~\ref{figure1}(a)] with a typical configuration shown in Fig.~\ref{figure1}(b).

Importantly, solid-fluid coexistence was observed to be a stable configuration for a range of $\phi$ \textit{at all values of activity} -- including those in which MIPS is observed.
At small activities, where particle motion is nearly reversible~\cite{Note2}, systems approach the well-established hard-sphere coexistence densities~\cite{Hoover1968, Pusey2009} of $\phi_{\rm fluid}=0.494$ and $\phi_{\rm solid}=0.545$. At $\ell_0/\sigma = 0.05$, for instance, we find $\phi_{\rm fluid}=0.52$ and $\phi_{\rm solid}=0.58$.
With increasing activity, we observe a rapid departure from this reversible limit; coexisting densities of both phases increase markedly.
The solid packing fraction quickly approaches the close-packed volume fraction $\phi_{\rm cp} \approx 0.74$ and remains near this value for all $\ell_0/\sigma \ge 5$.

In contrast to the solid density, the activity dependence of the fluid is nonmonotonic and defines some of the central features of the phase diagram.
As the activity is increased from zero, the fluid density rapidly increases to a volume fraction of $\approx 0.59$ (at $\ell_0/\sigma = 5$), then \textit{decreases} upon reaching the critical activity for MIPS.
The fluid density continues to decrease with activity until intersecting the MIPS binodal ($\ell_0/\sigma\approx 21.25$) slightly above the critical point.
The intersection of these coexistence curves results in an \textit{active triple point}~\footnote{Three-phase coexistence occurs at a single point
  in the pressure-activity plane but for a continuous range of $\phi$ in the
  density-activity plane.} where gas, liquid, and solid phases can coexist at the densities marked in Fig.~\ref{figure1}(a). 

Above the triple activity, the fluid that coexists with the solid phase has a density that is strictly less than the MIPS gas-phase density.
As a result, above the three-phase coexistence line, the liquid-gas binodal is entirely engulfed by the solid-fluid coexistence boundary [see Fig.~\ref{figure1}(a)].
In an equilibrium context, encapsulation of the liquid-gas binodal by the crystal-fluid phase boundary is a familiar and generic feature of simple substances below their triple temperature~\cite{ReinTenWolde1997, Dijkstra1999, Haxton2015}. 
Equilibrium requirements that free energy be convex and extensive further guarantee that the phase boundary with more extreme densities (typically crystal-fluid) corresponds to the more stable coexistence.
Leveraging the tools of large deviation theory~\cite{Touchette2009}, a similar conclusion can be drawn even for systems out of equilibrium~\cite{Note2}. 
For our ABPs at activities above $\ell_0/\sigma\approx 21.25$, states of liquid-gas coexistence should therefore crystallize irreversibly.
We observe that systems above the triple point and within the MIPS binodal can nevertheless persist for very long times in a state of liquid-gas coexistence. 
We now aim to verify that these states are globally unstable.

\textit{Homogeneous nucleation and stability.--} Despite recent progress in the development of importance sampling techniques for nonequilibrium systems~\cite{Giardina2011, Klymko2018, Ray2018, Das2019a, Whitelam2020, Ray2020, Helms2020, Oakes2020, Kuznets-Speck2021, Rose2021}, the ability to comprehensively survey the phase behavior of many-particle active systems~\cite{Nemoto2019, GrandPre2018, GrandPre2021, Keta2021} remains limited.
In the absence of these tools, we make an appeal to two-state rate theory to identify the relative stability of the two coexistence scenarios. 
Observing one form of coexistence (e.g.,~liquid-gas) spontaneously transition to the other (e.g.,~solid-fluid), and failing to observe the reverse transition, would provide compelling evidence for the global stability of the latter coexistence scenario (and, naturally, the metastability of the former).
However, long simulations at many such state points reveal no transitions. 
For example, Fig.~\ref{figure1}(e) shows the time evolution of the global order parameter $Q_{12} = \langle q_{12} \rangle$ (the particle-averaged $q_{12}$) at ($\ell_0/\sigma = 50$, $\phi=0.5$).
This points to the looming larger question: Can we observe the \textit{unbiased} nucleation of an active crystal from a disordered fluid?
We therefore turn to understanding the general forces that sculpt the crystal nucleation landscape and their dependencies on the state parameters ($\ell_0/\sigma$, $\phi$).

Figure~\ref{figure2} surveys the crystal nucleation landscape at dense packing fractions \textit{outside} of the liquid-gas binodal ($\phi > 0.61$ for $\ell_0/\sigma = 50$). 
In this region of the phase diagram, solid-fluid coexistence is the unambiguously stable system state.
We prepare these metastable high-density fluids by the isotropic compression of less-dense fluids~\cite{Note2}.
A disordered fluid at $\phi=0.635$ is observed to remain a liquid on all accessible timescales.
Fluids at $\phi \ge 0.64$, by contrast, readily nucleate a tightly packed active crystal (fcc), which grows into a single ordered domain that coexists with a fluid (gas) bubble [see Fig.~\ref{figure2}(a)]. 
The crystal symmetry and coexisting densities are consistent with those obtained from our crystal seeding procedure.
Crystal nucleation remains facile up to $\phi = 0.65$ (near maximal packing~\cite{Torquato2000}), the limiting density at which a hard-sphere fluid can still relax.

The remarkably narrow window of density ($0.64 \le \phi \le 0.65$) where active crystal growth can be observed makes evident why the 3D active order-disorder transition has, to our knowledge, previously eluded observation.
That this nucleation window occurs near maximal packing can be understood from general ideas of classical nucleation theory, which has successfully described the nucleation of 2D active liquids~\cite{Richard2016, Redner2016, Levis2017}. 
In this framework, the characteristic crystal nucleation rate should be governed by the product of the inverse fluid relaxation time $\tau_{\rm fluid}^{-1}$ and the probability $P_{\rm CN}$ of forming the critical nucleus in the course of spontaneous fluctuations~\footnote{In equilibrium, this probability follows from the critical
  nucleus free energy barrier height (along the appropriate reaction
  coordinate), e.g.,~$P_{\protect \text {CN}}\sim \protect \qopname \relax
  o{exp}[-\Delta G^{\protect \dagger }/k_{\protect B}T]$.}.

High densities are generally considered inhospitable for nucleation, since fluids typically vitrify near maximal packing, i.e.,~$\tau_{\rm fluid}$ diverges. 
Highly active fluids, however, exhibit no sign of glassy dynamics 
up to a density of (at least~\footnote{We cannot characterize the dynamics of the active fluids at $\phi =0.640$ and $0.645$, as the nucleation time is comparable to the fluid relaxation time. The short lifetime of these fluids also precludes the inclusion of these densities in the structural analysis presented in Figs.~\ref{figure2}(b) and \ref{figure2}(c).}) $\phi=0.635$, as evidenced by the self-component of the dynamic structure factor [Fig.~\ref{figure2}(b)]. 
Significant arrest only occurs upon reaching the geometrically frustrated maximal random limit, consistent with the emerging active glass literature~\cite{Ni2013, Berthier2014, Flenner2016, Berthier2017, Berthier2019, Nandi2018, Takatori2020}.

In the absence of vitrification, dense liquids can be favorable for nucleation, since they promote fluctuations that feature solidlike local density.
We quantify this enhancement of $P_{\rm CN}$ by calculating the probability $P_v(N)$ to observe $N$ particles in a spherical probe volume $v$ of diameter $\mathcal{D}_p$. 
Much like hard spheres at equilibrium~\cite{Crooks1997}, the distribution is Gaussian for many standard deviations, even for large densities and relatively small probe diameters [see Fig.~\ref{figure2}(c)].
For $v$ comparable in size to a plausible critical nucleus ($\mathcal{D}_p = 6\sigma$), solidlike local densities are highly atypical at $\ell_0/\sigma=50$ for fluids at all densities we have studied, shown in Fig.~\ref{figure2}(d). 
For $\phi \le 0.635$, such extreme local density fluctuations are so unlikely as to be unobserved in our long simulations. Near $\phi=0.65$, they become discernible (while still rare), consistent with our observations of successful crystal nucleation.

\begin{figure}
	\centering
	\includegraphics[width=.475\textwidth]{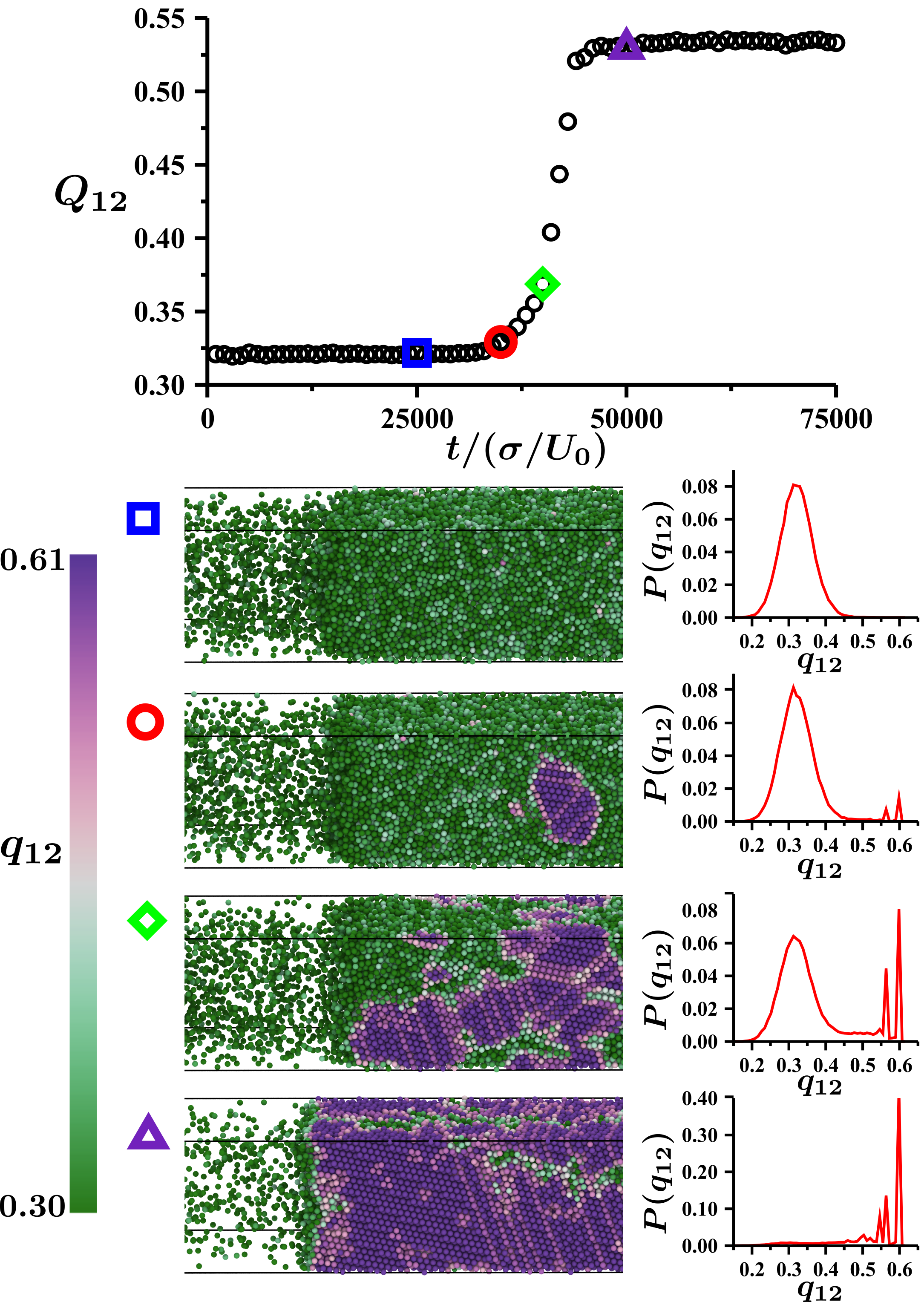}
	\caption{\protect\small{{
	Spontaneous transition from liquid-gas to solid-fluid coexistence $(\ell_0/\sigma = 500, \phi = 0.5)$. Time evolution of $Q_{12}$, indicating four frames whose structures are (partially) rendered alongside the distributions of $q_{12}$ for each configuration.}}}
	\label{figure3}
\end{figure}

\textit{Discussion and conclusions.--} The near close-packed density of active crystals severely restricts the region of the phase diagram where crystallization can feasibly be observed in simulations, in stark contrast to the broad range of conditions in which MIPS is readily observable (via nucleation or spinodal decomposition). 
Sufficiently close-packed local fluctuations occur with non-negligible probability only in fluids that are almost maximally packed. 
Direct observation of transitions from liquid to solid, such as in Fig.~\ref{figure3}, is thus feasible at very high activity, where the liquid phase is extremely dense. 
At low activities, the MIPS liquid-phase density is simply too low to nucleate an active crystal on accessible timescales, as exemplified by the long trajectory of Fig.~\ref{figure1}(e) with liquid density $\phi \approx 0.61$. In this low-activity case we lack the direct evidence of spontaneous transitions to judge the metastability of liquid-gas coexistence. 
Based on theoretical considerations~\cite{Note2}, however, the densities of coexisting phases we observe in simulations constitute strong indirect evidence to this effect. 
We therefore conclude that MIPS is in fact \textit{metastable} above the triple point activity. 
Consequently, liquid-gas coexistence is only the globally stable state in the narrow interval~\footnote{We emphasize that in the athermal hard-sphere limit, no adjustable parameters are available to tune the relative location of the critical and triple points -- the close proximity of these points (and the resulting stability implications) is an intrinsic property of the system.} between the critical and triple points [see Fig.~\ref{figure1}(a) inset and videos in Supplemental Material~\cite{Note2}].

The phase diagram presented in this work bears a striking resemblance to the phase diagram of traditional equilibrium molecular or colloidal systems with short-ranged attractions~\cite{ReinTenWolde1997, Dijkstra1999, Haxton2015}. 
However, attempting to directly equate activity to an ``effective attraction" has proven to be difficult~\cite{Farage2015, Rein2016, Turci2021}.
We therefore anticipate that 3D active hard spheres will serve as an important system to generalize the equilibrium arguments used in the construction of solid-fluid phase boundaries (and triple points) to nonequilibrium systems.
Moreover, additional examination of active phase behavior in 3D may prove insightful for further understanding the role of dimensionality in the rich phase behavior (such as ``bubbly liquids''~\cite{Tjhung2018, Shi2020, Caporusso2020}) reported in 2D. 
Finally, while active freezing has primarily been experimentally interrogated in 2D~\cite{Briand2016, Briand2018, Huang2020}, we hope that our study will aid in guiding ongoing efforts~\cite{Sakai2020} to realize 3D active crystals.

\begin{acknowledgments}
A.K.O. acknowledges support from the Schmidt Science Fellowship in partnership with the Rhodes Trust.
P.L.G. was supported by the U.S. Department of Energy, Office of Basic Energy Sciences, through the Chemical Sciences Division (CSD) of Lawrence Berkeley National Laboratory (LBNL), under Contract No. DE-AC02-05CH11231.
This research used the Savio computational cluster resource provided by the Berkeley Research Computing program. 
We gratefully acknowledge the support of the NVIDIA Corporation for the donation of the Titan V GPU used to carry out part of this work.
\end{acknowledgments}

\end{document}


\title{Supplemental Material -- Phase Diagram of Active Brownian Spheres: Crystallization and the Metastability of Motility-Induced Phase Separation}

\author{Ahmad K. Omar}
\email{aomar@berkeley.edu}
\affiliation{Department of Materials Science and Engineering, University of California, Berkeley, California 94720, USA}
\affiliation{Department of Chemistry, University of California, Berkeley, California 94720, USA}
\author{Katherine Klymko}
\affiliation{Computational Research Division, Lawrence Berkeley National Laboratory, Berkeley, California 94720, USA}
\affiliation{NERSC, Lawrence Berkeley National Laboratory, Berkeley, California 94720, USA}
\author{Trevor GrandPre}
\affiliation{Department of Physics, University of California, Berkeley, California 94720, USA}
\author{Phillip L. Geissler}
\email{geissler@berkeley.edu}
\affiliation{Department of Chemistry, University of California, Berkeley, California 94720, USA}
\affiliation{Chemical Sciences Division, Lawrence Berkeley National Laboratory, Berkeley, California 94720, USA}

\maketitle

\section{Supplemental Videos}
The videos included in the Supplemental Material are intended to serve as representative examples of the states/transitions listed below. For reference to the expected globally stable states as a function of ($\ell_0/\sigma$, $\phi$), see the phase diagram provided in the main text [Fig.~1(a)]. In all videos, the captioned time is in units of $\sigma/U_0$. All videos are available at: \url{https://berkeley.box.com/s/ivwqz4ux1v9wva1sce4qzkpzitpewcyl}.
\begin{enumerate}
    \item \textbf{Metastable Fluid} \newline metastable\_fluid\_no\_nucleation.mp4 \newline ($\ell_0/\sigma = 50$, $\phi=0.62$) \newline Illustrates the absence of nucleation events in a metastable fluid.
    \item \textbf{Crystal Nucleation in a Metastable Fluid} \newline metastable\_fluid\_nucleation.mp4 \newline ($\ell_0/\sigma = 50$, $\phi=0.64$) \newline Demonstrates facile nucleation as maximal fluid packing is approached.
    \item \textbf{Liquid-gas to Solid-fluid Coexistence in Metastable MIPS Region} \newline liquid-gas\_to\_solid-fluid.mp4 \newline ($\ell_0/\sigma = 500$, $\phi=0.5$) \newline Spontaneous crystallization within a liquid (coexisting with a gas), demonstrating the global stability of solid-fluid coexistence [see Fig.~3 in the main text]. 
    \item \textbf{Solid-fluid to Liquid-gas Coexistence in Stable MIPS Region} \newline solid-fluid\_to\_liquid-gas.mp4 \newline ($\ell_0/\sigma = 20$, $\phi=0.5$) \newline Melting of a crystal in the narrow area of the phase diagram where liquid-gas coexistence is the globally stable state.
    \item \textbf{Stability of Solid-Fluid Coexistence} \newline solid-fluid\_stability\_above\_triple\_point.mp4 \newline ($\ell_0/\sigma = 24$, $\phi=0.5$) \newline Solid-fluid coexistence is stable for long-times within the MIPS binodal and above the triple point activity. Shortly below the triple point [see Video~4], the solid will melt.
\end{enumerate}

\section{The Equilibrium Limit of Active Brownian Spheres}
It is well known that active particles (both the active Brownian particles examined in this work and those that follow the active Ornstein-Uhlenbeck process) behave as particles with equilibrium statistics in the limit of a vanishing reorientation time $\tau_R \rightarrow 0$~\cite{Fodor2016, Speck2016, Mandal2017}. 
Here, we provide a simple argument for the reversible limit by demonstrating that in the limit of vanishing reorientation time, the statistics of the active force are indistinguishable from that of the thermal Brownian force acting on equilibrium particles. 
In other words, the active force satisfies the fluctuation dissipation theorem (FDT) in this limit. 

Neglecting rotational inertia and alignment interactions between particle orientations results in an independent first-order stochastic differential equation for each particle's orientation $\mathbf{q}(t)$. 
The resulting orientation time autocorrelation function follows as:
\begin{equation}
\label{q_correlations}
    \langle{\mathbf{q}(t)\mathbf{q}(0)}\rangle = e^{-t(d-1)/\tau_R}\mathbf{I}/d, 
\end{equation}
where $d$ is the dimensionality of the system and $\mathbf{I}$ is the identity tensor. 
The memoryless drag force used in this work ($-\zeta  \mathbf{\dot{x}}$) constrains the functional form of any additional nonconservative forces (in this case, the active force) if FDT is to be rigorously satisfied. 
In addition to having zero mean, for the active force to satisfy FDT the variance must take the following form:
\begin{equation}
    \langle{\mathbf{F^A}(t)\mathbf{F^A}(0)}\rangle = 2T^{\text{eff}}\zeta \delta(t) \mathbf{I}, 
\end{equation}
where $T^{\text{eff}}$ has units of energy. 
To satisfy FDT, the active force must be delta-correlated in time. 
For ABP's, the variance of the active force is:
\begin{equation}
\label{activeFDT}
    \langle{\mathbf{F^A}(t)\mathbf{F^A}(0)}\rangle = \zeta^2 U_0^2 e^{-t(d-1)/\tau_R}\mathbf{I}/d.
\end{equation}

Equation~\eqref{activeFDT} clearly demonstrates that while the active force generally violates FDT, there is a well-defined reversible limit. 
Multiplying and dividing eq.~\eqref{activeFDT} by $\tau_R/(d-1)$ and taking the limit as $\tau_R \rightarrow 0^+$ results in:
\begin{equation}
    \langle{\mathbf{F^A}(t)\mathbf{F^A}(0)}\rangle = 2T^{\text{eff}}\zeta \delta(t) \mathbf{I},
\end{equation}
with
\begin{equation}
    T^{\text{eff}} = \frac{\zeta U^2_0 \tau_R}{d(d-1)}.
\end{equation}
This limit is the \textit{only scenario} in which the active force can be delta-correlated in time and rigorously satisfy FDT. 

For a given physical system, this limit is more appropriately expressed by comparing $\tau_R$ with the other system timescales. 
For repulsive active particles, the only other natural timescale is the convection time: the ratio of the particle size to the intrinsic active speed $\sigma /U_0$.
The limit for effective equilibrium of repulsive active particles is therefore $\ell_0 /\sigma \rightarrow 0$ as stated in the introduction in the main text. 
Note that while the effective temperature will also become vanishingly small in this limit (if the active force magnitude $\zeta U_0$ is held fixed), the lack of a competing energy scale renders the precise value of temperature inconsequential for the stationary distribution of hard particles with reversible dynamics.

\section{Stability Consequences of a Nonequilibrium Triple Point}

\begin{figure}
	\centering
	\includegraphics[width=.475\textwidth]{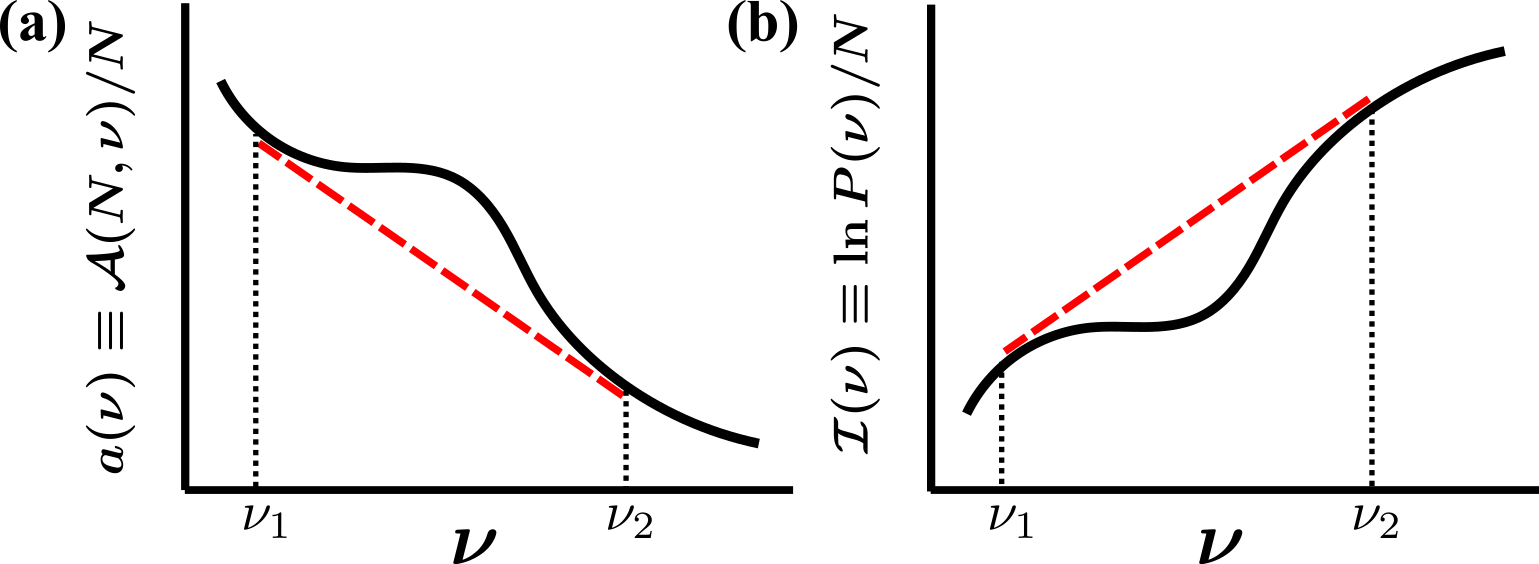}
	\caption{\protect\small{{In equilibrium, the coexisting densities can be found by (a) performing a common tangent construction (in the per-particle volume $\nu$ plane) on the per-particle Helmholtz free energy $a(\nu)$. The generalization of this procedure is (b) a common tangent construction (also in $\nu$) on the rate function $\mathcal{I}(\nu)$ that characterizes volume fluctuations in the absence of external constraints with $\mathcal{I}(\nu) = -a(\nu)/k_BT$ in equilibrium.}}}
	\label{sm_figure1}
\end{figure}

\begin{figure*}[!t]
	\centering
	\includegraphics[width=.95\textwidth]{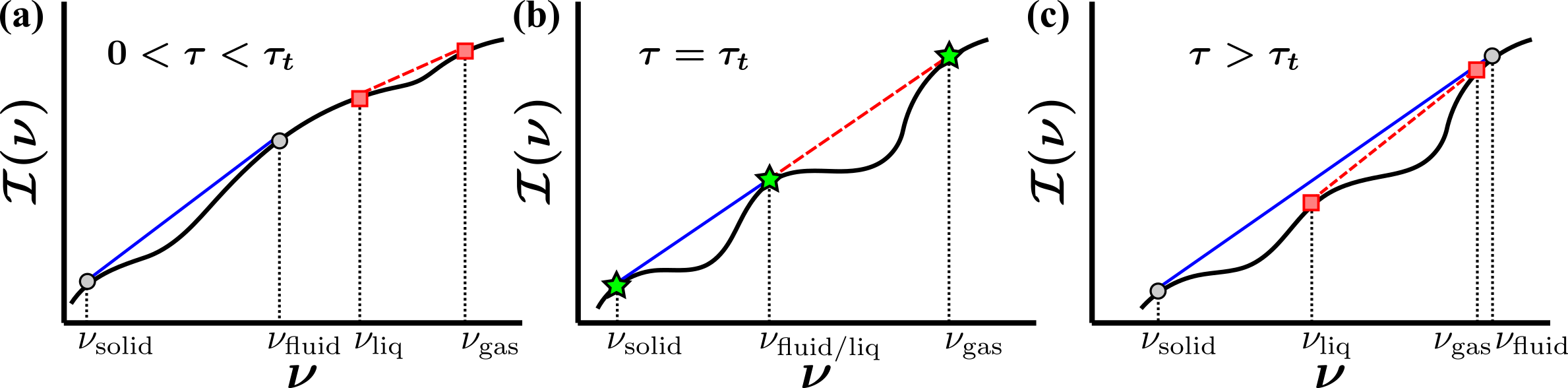}
	\caption{\protect\small{{Schematic of the rate function (a) below (while above the critical point), (b) at and (c) above the triple point $\tau_t$. Straight solid and dashed lines are the result of common tangent constructions for solid-fluid and liquid-gas coexistence, respectively.}}}
	\label{sm_figure2}
\end{figure*}

Our work raises important questions concerning the nature of a nonequilibrium triple point, the most natural being: is MIPS metastable above the triple activity?  
Explicitly observing a transition from liquid-gas to solid-fluid coexistence for activities near and modestly above the triple point would require prohibitively long trajectories (lacking tools of importance sampling that would enable such a task for equilibrium systems). 
Our study makes clear that it is only when the liquid-phase density exceeds a value of $\phi=0.635$ that nucleation can be observed on feasible time scales. 
The range of activity that permits direct confirmation of metastability is therefore quite limited. 

However, shared features between the equilibrium phase diagram of simple materials and our active system suggest that triple points may have generic features, whether a system is in equilibrium or otherwise.
Much like an equilibrium fluid below its triple temperature, we observe the complete encapsulation of the liquid-gas binodal by the crystal-fluid phase boundary above the triple activity [see Fig.~1(a) in the main text]. (By encapsulation, we mean that the liquid-gas coexistence densities both lie in the range between solid-fluid coexistence densities at the same activity.)
In equilibrium, this relative positioning between the phase boundaries also coincides with the metastability of liquid-gas coexistence for \textit{all} temperatures below the triple temperature.
Our aim here is to establish whether these features hold for nonequilibrium systems.

Consider a collection of $N$ particles enclosed in a region of fluctuating \textit{unconstrained} volume $V$. 
Here, the system volume should be thought of as an internal degree of freedom that is not externally constrained or regulated by an external pressure. 
We will first examine the stability of spatially uniform states, which have finite correlation length. 
We describe the conditions for which nonuniform states (e.g.,~phase separation and crystallization) are more probable when the system volume is held fixed.
From these considerations, we establish that the relative stabilities of competing coexistence scenarios can be simply deduced from their relative coexistence densities regardless of whether a system is in thermodynamic equilibrium or not. 

Let us first consider the case of an equilibrium system before proceeding to generalize our arguments. 
The relevant thermodynamic potential for the conditions described above is the Helmholtz free energy $\mathcal{A}(N, \nu)/k_BT$ which depends on $N$ (or $V$) and the per-particle volume $\nu\equiv V/N$.
Thermodynamic stability dictates that the intensive (per particle) free energy density $a(\nu) \equiv \mathcal{A}(N,\nu)/N$ must be convex with respect to all unconstrained degrees of freedom, which in this case is simply $\nu$.\footnote{Additionally, $a(\nu)$ is typically a monotonically decreasing function of $\nu$, as most physical systems exert a positive pressure $\mathcal{P} = -\partial a(\nu)/ \partial \nu$. Nonmonotonicities in $a(\nu)$ (i.e.,~negative pressures) can arise due to the condition of spatial uniformity (see, for example, the Van der Waals pressure and free energy density when well below the critical temperature).}
Concavity of the free energy density in $\nu$ results in the spontaneous reduction of free energy by phase separation into domains of coexisting particle volumes $\nu_1$ and $\nu_2$ when the system volume is held fixed. 
These coexisting densities are found by minimizing the total system free energy, which is taken to be the sum of each phase's free energy i.e.,~$\mathcal{A} = \mathcal{A}(N_1, \nu_1) + \mathcal{A}(N_2, \nu_2)$ (which neglects interfacial effects), with respect to the particle number/volume of either phase. 
This results in a common tangent construction on the free energy density\footnote{Defining the free energy density on a per particle basis results in a common tangent construction in $\nu$, while defining it on a per volume basis would result in a common tangent construction in density, $\nu^{-1}$.}, shown schematically in Fig.~\ref{sm_figure1}(a).
The free energy density at a particle volume $v_0$ that lies between $\nu_1$ and $\nu_2$ is simply a linear interpolation between the free energy densities of the coexisting states, with the relative fraction of each phase provided by the lever rule resulting from mass conservation.  

In equilibrium, the free energy provides the statistical weight for fluctuations in the volume per particle $\nu$, with:
\begin{equation}
    P(\nu) \propto \exp[-\mathcal{A}(N,\nu)/k_BT].
\end{equation}
However, this form of the probability, the large deviation form~\cite{Touchette2009}, extends beyond equilibrium.
The extensivity of volume combined with finite particle correlation length (resulting in statistically independent subvolumes for large system volumes) must result in the probability of observing the system with per-particle volume $\nu$ having a large deviation form~\cite{Touchette2009}:
\begin{equation}
\label{eq:ldt}
    P(\nu) = \exp[N\mathcal{I}(\nu)],
\end{equation}
where $\mathcal{I}(\nu)$ is the large deviation rate function that solely depends on $\nu$.
In equilibrium, we identify $\mathcal{A}(N,\nu)/k_BT = -N\mathcal{I}(\nu)$.

We can now recast the thermodynamic arguments used to predict phase coexistence using the more general language provided by large deviation theory. 
We assume that the probability for a state of coexistence has the form $\ln P(\nu) = N_1I(\nu_1) + N_2I(\nu_2)$, which simply amounts to neglecting interfacial effects and assuming the phases are large enough that each follows large deviation statistics. 
In regions of particle volume in which $\mathcal{I}(\nu)$ for spatially uniform systems is convex, the system can spontaneously evolve to a more probable state by phase separating [see Fig.~\ref{sm_figure1}(b)] when the system volume is held fixed. 
Maximizing the system probability again results in a common tangent construction on $\mathcal{I}(\nu)$. 

To summarize, regardless of whether a system is in equilibrium or not, the large deviation form [eq.~\eqref{eq:ldt}] is anticipated to characterize volume fluctuations for system sizes larger than the particle correlation length. 
The rate function describing these fluctuations for spatially uniform states must be concave, or the system will phase separate if the volume is held fixed.
The coexisting densities are then found by maximizing probability with respect to the unconstrained internal variables, the particle number/volume in the phases.
We emphasize that determining the rate function that describes density fluctuations for active systems (explored in Ref.~\cite{Chakraborti2016}) remains an ongoing challenge due to the absence of robust importance sampling techniques for driven systems. 
Our analysis here is schematic in nature, and only requires an understanding of the basic features of $\mathcal{I}(\nu)$.

Equipped with the language to understand equilibrium and nonequilibrium coexistence on the same footing, we can now finally address the question raised at the start of this section: is liquid-gas coexistence metastable above a nonequilibrium triple point?
We consider the rate function of a generic system that has two forms of coexistence, solid-fluid and liquid-gas and a triple point at $\tau = \tau_t$. Here, $\tau$ is a generalized critical parameter, defined so that (a) $\tau=0$ locates the liquid-gas critical point, and (b) three-phase coexistence occurs at a positive value $\tau_t >0$  (e.g.,~$\tau = 1 - T/T_c$ for a simple equilibrium substance or $\tau = \ell_0/\ell_c - 1$ for our active system). 
For $0<\tau<\tau_t$, the rate function $\mathcal{I}(\nu)$ has two distinct intervals of convexity separated by a stable (concave) region, as shown schematically in Fig.~\ref{sm_figure2}(a). 
The convexity at low $\nu$ is resolved through solid-fluid coexistence, while that at larger $\nu$ results in liquid-gas coexistence.
As $\tau$ approaches the triple point, the observed broadening of each phase boundary is a result of the widening convex intervals. 
At the triple point, the common tangent construction on each convex region results in identical liquid/fluid densities, with a single line connecting all three coexisting phases [see Fig.~\ref{sm_figure2}(b)]. 

For $\tau > \tau_t$, the two distinct convex regions continue to widen (with the intermediate region of stability narrowing) resulting in a jump of the crystallization fluid density to a density that is strictly less than the gas-phase density (i.e.,~$\nu_{\rm fluid} > \nu_{\rm gas}$) of liquid-gas coexistence as shown in Fig.~\ref{sm_figure2}(c).
All that is required for this to hold is that the two common tangent constructions generate phase boundaries with overlapping densities and that the liquid-phase density is less than that of the solid (i.e.,~$\nu_{\rm liq} > \nu_{\rm solid}$). 
The encapsulation of the liquid-gas phase boundary by crystallization above the triple point is thus a generic feature of all triple points. 
Coincident with this encapsulation is that the solid-fluid coexistence line \textit{always} lies above that of liquid-gas coexistence, i.e.,~its probability is higher. 
As a result, \textit{crystallization must be the globally stable form of coexistence above the triple point for both equilibrium and nonequilibrium systems.}

In fact, the argument above applies much more broadly: if multiple coexistence scenarios compete (i.e.,~there exist values of density where multiple forms of coexistence are possible, as is the case for $\tau > \tau_t$), one form of coexistence \textit{must} have more extreme densities than the other(s) and that form \textit{must} be the globally stable state. 
This conclusion, rooted in an analysis that is applicable to both equilibrium and nonequilibrium systems, suggests that questions of global stability can be resolved by simple examination of the coexisting densities.
This is particularly powerful for nonequilibrium systems, where robust methods of importance sampling have made questions of global stability particularly challenging to address.
These arguments are anticipated to hold more generally, including for order parameters beyond density. 

Observing the explicit transition between two forms of coexistence at a state point remains the gold standard for establishing relative stabilities at that point in the phase diagram. 
But explicit verification is computationally prohibitive for our system over a large range of activity. 
Instead, we (1) established the physics that underpin the nucleation barrier to understand which state points we can explicitly verify (high activity), (2) verified that the conclusions from the nucleation study are borne out [Fig.~3 in the main text], and (3) used general statistical mechanical considerations to argue that the entire region of liquid-gas coexistence must be metastable above the triple point activity. 
We believe this collection of evidence establishes the metastability of MIPS above the triple point activity. 

\section{Additional Simulation Details}
As stated in the main text, the stiffness parameter $\mathcal{S} = \varepsilon/(\zeta U_0 \sigma)$ plays a determining role in recovering hard-sphere packing statistics.
This can be immediately appreciated by noting that the force between a pair of particles can be expressed as $\mathbf{F}_{ij}^{\mathbf{C}}(\mathbf{x}_{ij}) = -\boldsymbol{\nabla}\varepsilon u^\text{WCA}(r; \sigma)$ where $\mathbf{x}_{ij}$ is the particle separation (with magnitude $r = |\mathbf{x}_{ij}|$) and the reduced potential $u^\text{WCA}$ has the following form~\cite{Weeks1971}:
\begin{equation}
u^\text{WCA}(r; \sigma) =
 \begin{cases}
   4\left[\left(\frac{\sigma}{r}\right)^{12} - \left(\frac{\sigma}{r}\right)^{6}\right] + 1, & r \le 2^{1/6}\sigma \\
    0, & r > 2^{1/6}\sigma.  \\
    \end{cases}
\end{equation}
Upon selecting $\zeta U_0$ and $\sigma$ as the respective units of force and length, the resulting dimensionless force $\overline{\mathbf{F}_{ij}^{\mathbf{C}}}(\overline{\mathbf{x}}_{ij}; \mathcal{S}) = -\mathcal{S}\overline{\boldsymbol{\nabla}} u^\text{WCA}(\overline{r})$ (where $\overline{\boldsymbol{\nabla}} = \sigma \boldsymbol{\nabla}$ is the dimensionless gradient operator) is entirely characterized by the stiffness $\mathcal{S}$.
A choice of $\mathcal{S} = 50$ ensures that the active force cannot generate overlaps within a pair separation $d$ of $d/(2^{1/6}\sigma) = 0.9997$, allowing us to achieve effective hard-sphere statistics provided a suitable timestep is chosen.
Integrating our equation-of-motion [eq.~(1) in the main text] with a timestep of $5\times10^{-5}~\sigma/U_0$ (using the GPU-enabled \texttt{HOOMD-blue} software~\cite{Anderson2020}) is found to be sufficient to ensure minimal particle overlap. 

To quantify bond-orientational order in our system, we compute the per-particle Steinhardt-Nelson-Ronchetti order parameter~\cite{Steinhardt1983}:
\begin{equation}
\label{eq:localq12pt1}
q_l(i) = \biggl (\frac{4\pi}{2l+1} \sum_{m=-l}^{l}|\langle Y_{lm}\rangle|^2 \biggr)^{1/2},
\end{equation}
where $\langle Y_{lm} \rangle$ is the average spherical harmonics of the bond angles formed between particle $i$ and its nearest neighbors. 
By setting $l=12$, we can readily identify particles in a perfect fcc arrangement as those with $q_{12} \approx 0.6$. 
We find that both liquid and gas phases are well described by a Gaussian distribution of $q_{12}$ centered at $q_{12}\approx 0.3$.
A useful measure for monitoring the kinetics of crystal growth is the global order parameter:
\begin{equation}
\label{eq:Q12}
Q_l = \frac{1}{N}\biggl\langle \sum_{i=1}^{N}q_{l}(i) \biggr\rangle,
\end{equation}
which is simply the average of $q_{l}$.

The distribution of the local (per particle) volume fraction is determined by computing the local Voronoi volume of each particle $V_{\text{Vor}}$.
The local volume fraction of an individual particle is then simply $\phi = \pi (2^{1/6} \sigma)^3/6V_{\text{Vor}}$. 
The location of the peaks of the distribution of $\phi$ obtained from this local definition provides liquid and gas-phase densities (far from the critical point) that are consistent with those obtained from slab simulations (see below).
The utility of computing this distribution is primarily in characterizing the crystal phase density as detailed below. 

\section{Construction of the Phase Diagram}

\subsection{The Liquid-gas Phase Boundary}

The liquid-gas phase boundary presented in the main text was found by conducting ``slab" simulations -- constant volume simulations with one box dimension larger than the other two ($L_z > L_x = L_y$). 
The asymmetric geometry results in a one-dimensional (along the long ($z$) axis) density profile that allows us to readily obtain the coexisting densities [see eq.~\eqref{eq:dprofile}].
Simulations were performed with a box aspect ratio of $L_z/L_x \approx 4$. 
The density $\phi$ was adjusted with activity to be approximately the geometric average of the anticipated liquid and gas-phase densities. 
This ensures that MIPS is spontaneous (which is only guaranteed within the spinodal, the limit of mechanical stability) and that we have a significant amount (by volume) of both phases.
Simulations at the following activity-dependent volume fractions were found to result in spontaneous phase separation: 
\begin{equation}
    \phi = 
    \begin{cases}
    0.35, & \ell_0 /\sigma \ge 30.0 \\
    0.45, & 19.5 \le \ell_0/\sigma < 30.0.  \\
    \end{cases}
\end{equation}

For nearly all simulations, the particles were initially arranged in a volume-spanning simple-cubic lattice. 
The relatively low concentrations prevents this rather arbitrary initial configuration from biasing the system to form ordered states. 
As activity is reduced and approaches the critical point ($\ell_0/\sigma \le 34$), we used the final configurations formed by systems with slightly higher activities as the initial condition for lower activities.
This allowed us to obtain the coexisting densities remarkably close to the critical point, a typically difficult task (in the absence of importance sampling) due to the reduced driving force for configurations with a single interface.
All simulations were conducted for a minimum duration of $1.5\times 10^{4}~\sigma/U_0$ and with $54900$ particles.

\begin{figure}
	\centering
	\includegraphics[width=.475\textwidth]{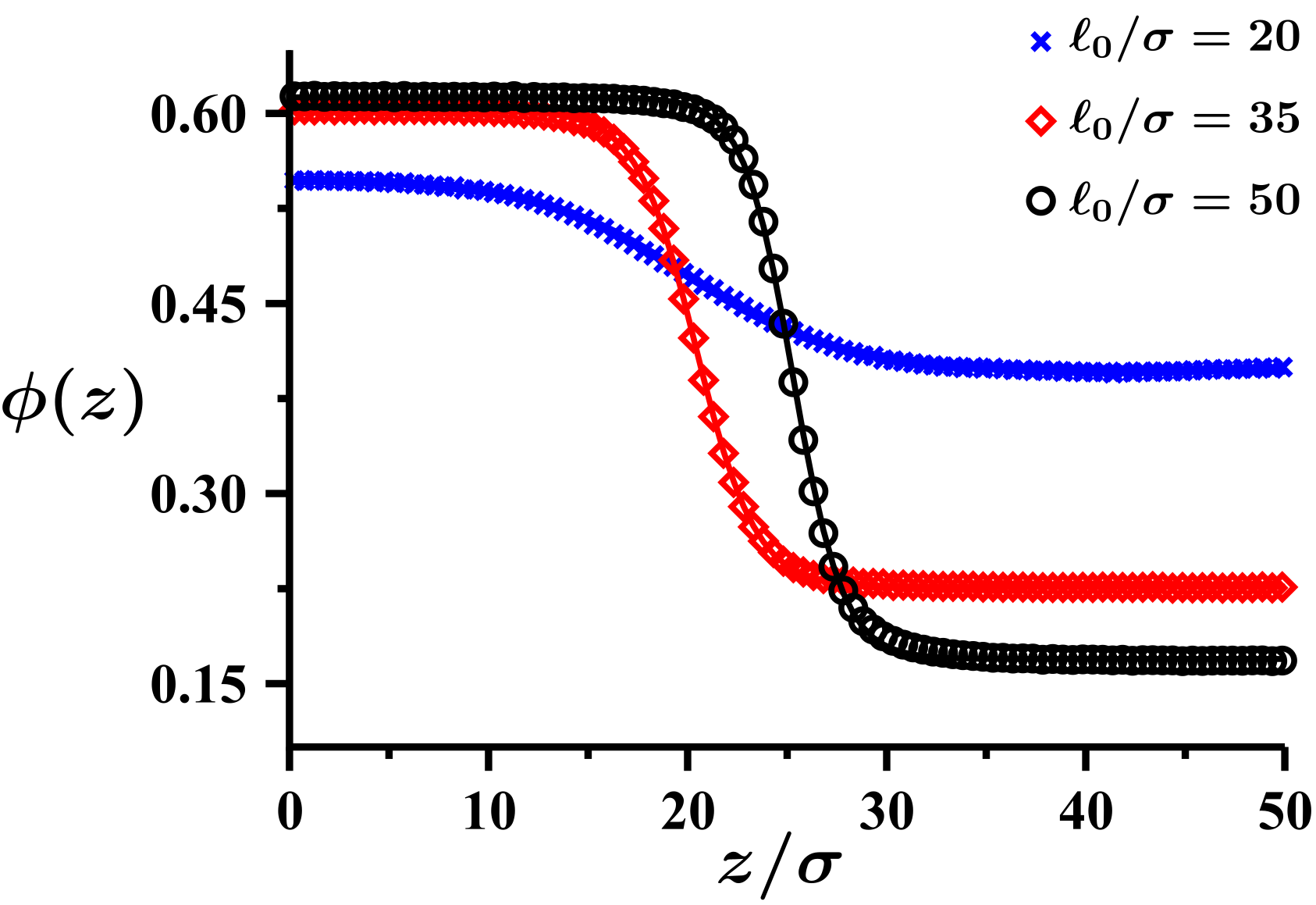}
	\caption{\protect\small{{Representative density profiles with symbols denoting simulation data while lines are fits using eq.~\eqref{eq:dprofile}}.}}
	\label{sm_figure3}
\end{figure}

Figure~\ref{sm_figure3} shows representative density profiles obtained from our simulations. 
These profiles were obtained by time averaging the density profiles of steady-state simulation trajectories while ensuring that the center-of-mass of the density profile is shifted to $z=0$.
For nearly all activities, these are extremely well-fit by a sigmoidal function:
\begin{equation}
\label{eq:dprofile}
\phi(z) = \frac{\phi_{\rm liq} - \phi_{\rm gas}}{2}\text{tanh}\left (\frac{z-z_0}{w} \right ) + \frac{\phi_{\rm liq} + \phi_{\rm gas}}{2},
\end{equation}
where $\phi_{\rm liq}$ and $\phi_{\rm gas}$ are the liquid and gas-phase densities (respectively) presented in our phase diagram, $w$ is the interfacial width and $z_0$ is a shift factor.
For large activities ($\ell_0 /\sigma \ge 250$), the sigmoidal fit was found to be less exact but remained acceptable.   

Finite size effects are anticipated to be important for all liquid-gas transitions near the critical point. 
However, for active systems, there is an additional region of the phase diagram in which finite size effects may prove influential.
At high activity, low density gas particles move relatively unobstructed, resulting in an effective run length that approximates large intrinsic run length $\ell_0$. 
As a result, the correlation length of a gas-phase particle's trajectory can be comparable to the domain size of the gas phase, the region bounded by gas-liquid interfaces as a result of periodic boundary conditions.

To isolate the role of system size on the high-activity region of the phase diagram, we conducted a limited set of simulations with $200000$ particles with an overall density of $\phi=0.32$ and a box aspect ratio of  $L_z/L_x \approx 5.4$.
The increased number of particles, reduced density and larger box aspect ratio act to dramatically increase the domain size of the gas phase in comparison to our primary set of simulation data.
For these large simulations, we initially prepare the system at a volume fraction $\phi = 0.55$ and perform a uniaxial elongation to reduce the density to our target value.
Such a procedure (which is similar to that of Ref.~\cite{Omar2020a}) is found to greatly reduce the likelihood of forming multiple liquid and gas domains that would require significantly longer simulations to coalesce.

\begin{figure}
	\centering
	\includegraphics[width=.475\textwidth]{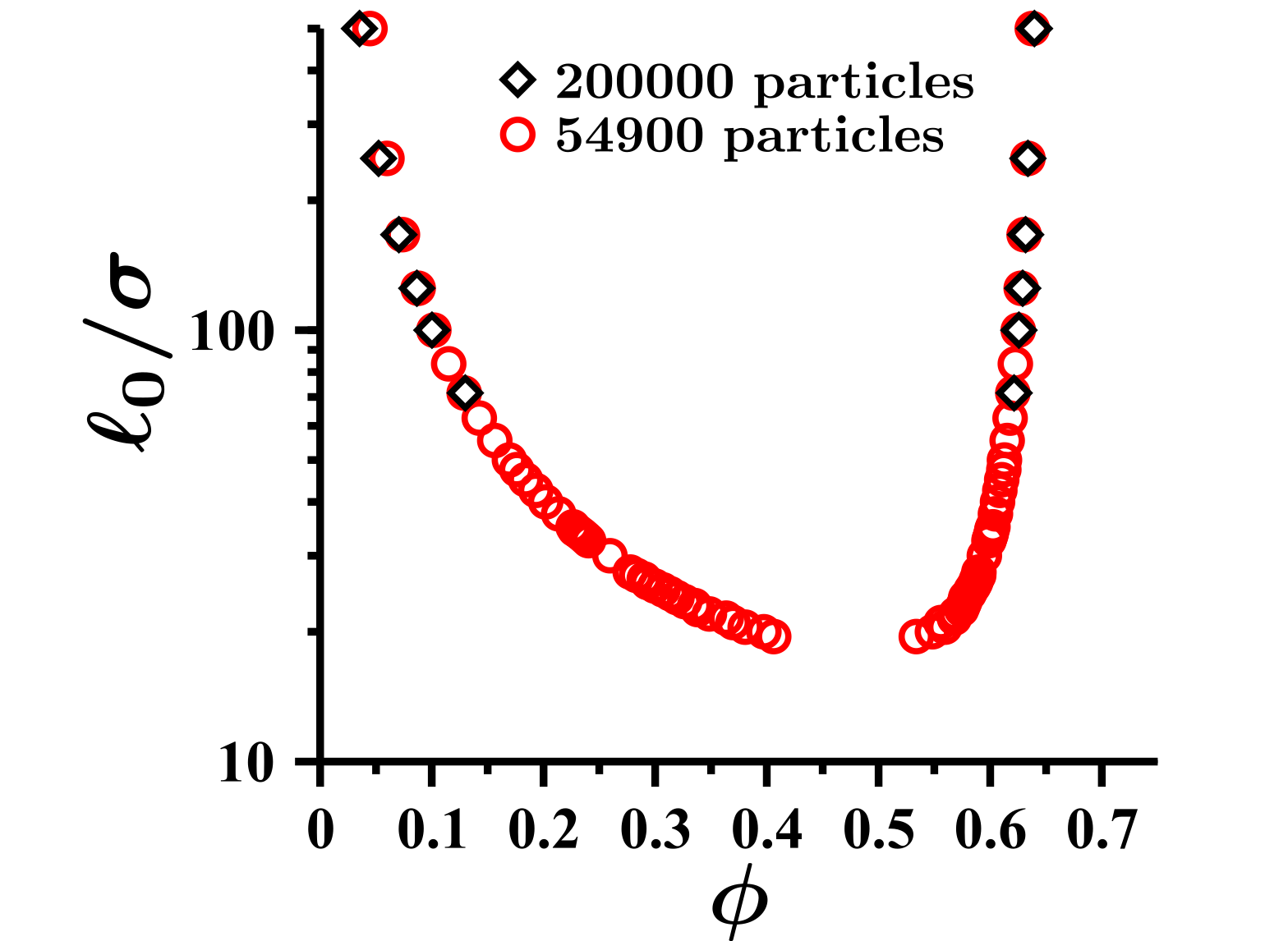}
	\caption{\protect\small{{Influence of system size on the high activity region of the MIPS binodal.}}}
	\label{sm_figure4}
\end{figure}

Figure~\ref{sm_figure4} illustrates near quantitative agreement between the coexisting densities resulting from our primary simulation data [Fig.~1(a) in the main text] and the larger simulations. 
Interestingly, the larger simulations do exhibit a small but notable reduction (relative to our primary simulations) of the gas-phase density with increasing activity.
Nevertheless, the good agreement between the two disparate system sizes suggests that activity-related finite size effects are not a significant concern for the conditions examined in this work.

\subsection{Critical Point}

The critical point of MIPS can be rigorously obtained by performing a system-size scaling and determining the activity in which the Binder parameter becomes scale invariant~\cite{Binder1981, Binder1987, Rovere1988, Rovere1990, Rovere1993}.
Such a procedure allows for the \textit{independent} determination of the critical activity and resulting critical exponents and has recently been applied to active systems~\cite{Siebert2018, Maggi2021, Dittrich2020, Adachi2020}. 
While conducting a full Binder analysis to determine the critical exponents of 3D MIPS is left for future work, we make use of a critical scaling ansatz to estimate the location of the critical point.
Defining the reduced activity as:
\begin{equation}
\tau \equiv \frac{\ell_0 - \ell_c}{\ell_c},
\end{equation}
and the order parameter to be:
\begin{equation}
\Delta \phi \equiv \phi_{\rm liq} - \phi_{\rm gas}, 
\end{equation}
(as is standard for a liquid-gas transition) we anticipate that:
\begin{equation}
\label{eq:tau}
\tau \propto \Delta \phi^\beta,
\end{equation}
where $\beta$ is a critical exponent.
This scaling is only anticipated to apply near the critical point (i.e.,~$\tau \ll 1$).
We therefore must self-consistently ensure that the data used to obtain a critical activity $\ell_c$ are sufficiently close to the produced critical activity.

The simulation data nearest to the critical point (i.e.,~the lowest activities where liquid-gas coexistence was observed) were fit to eq.~\eqref{eq:tau} by varying $\ell_c$, $\beta$, and a constant of proportionality. 
The optimal fit resulted in $\ell_c/\sigma \approx 18.5$ and $\beta=0.33$.
These parameters were subsequently used to obtain the critical density $\phi_c$ by fitting the liquid (and gas) phase densities to a critical scaling form $\phi_{\rm liq} -\phi_c \propto \tau^\beta$ consistent with eq.~\eqref{eq:tau}.  The resulting critical density was found to be $\phi_c \approx 0.483$.

\begin{figure}
	\centering
	\includegraphics[width=.475\textwidth]{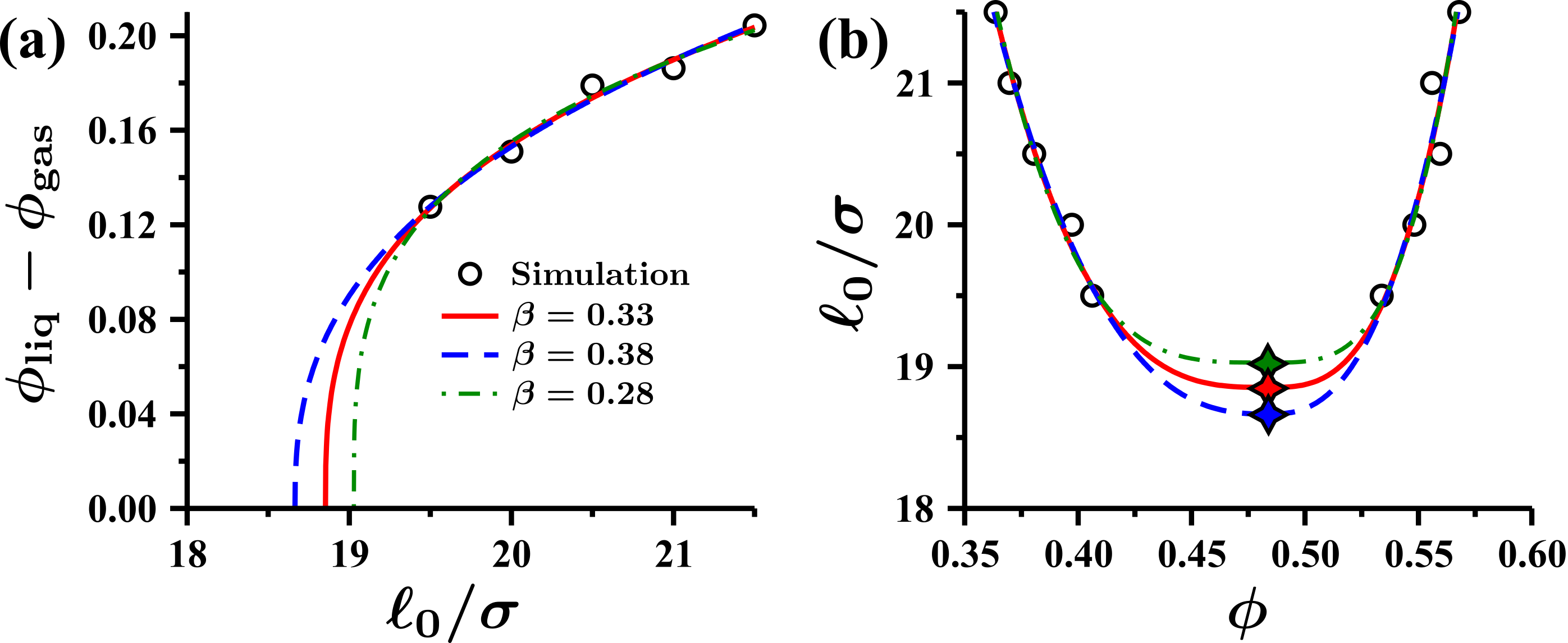}
	\caption{\protect\small{{Critical exponent $\beta$ dependence of (a) the liquid-gas order parameter as a function of activity and (b) the liquid-gas binodal. Only the simulation data shown was used in the critical scaling analysis.}}}
	\label{sm_figure5}
\end{figure}

While a critical exponent of $\beta \approx 0.33$ is in agreement with the 3D Ising universality class, simultaneously fitting $\ell_c$ and $\beta$ increases the uncertainty in our fit. 
Figure~\ref{sm_figure5} compares the results of our multiparameter fit with fits obtained by fixing $\beta$ to be $0.38$ and $0.28$.
For all of these values of $\beta$, the resulting order parameter and binodal fits agree well with the available simulation data (both the order parameter and binodal) with the primary difference in the fits being the location of the critical activity.
An independent determination of $\ell_c$ (e.g.,~by Binder analysis) will significantly reduce the uncertainty in the critical exponent. 

\subsection{The Solid-fluid Phase Boundary}
Our work makes clear that the conspicuous absence of 3D crystallization in the literature is due to the narrow range of the phase diagram where nucleation can be readily observed.
To establish the solid-fluid phase diagram, we implement a simple method that allows us to circumvent the kinetic barriers that may otherwise obstruct the determination of the coexisting densities. 
We initially arrange our particles in a \textit{perfect} fcc crystal at a volume fraction of $\phi = 0.7$. 
The box aspect ratio is initially $L_z/L_x \approx 2$ and our system consists of $54872$ particles.
With a stable crystal prepared, we begin to reduce the volume fraction of the system in increments of $0.02$ by elongating the $z$-axis at a rate of $0.1~U_0$.
At each volume fraction increment, the simulation is run for a relatively short duration of $500~\sigma/U_0$ to determine if solid-fluid coexistence is potentially a stable state.
We performed this ``sweep" until a volume fraction of $0.5$ is reached.

For each activity, we then independently determine the density at which solid-fluid coexistence appeared stable and contained a significant amount (by volume) of each phase.
Long-time simulations ($2.5\times10^{4}~\sigma/U_0$) were then run at these densities to verify the stability of the observed coexistence and extract the coexisting densities.
The following activity-dependent densities were found using this procedure: 
\begin{equation}
    \phi = 
    \begin{cases}
    0.54, & \ell_0/\sigma = 0.05 \\
    0.64, & \ell_0/\sigma = 0.5 \\
    0.66, & 0.5 < \ell_0 /\sigma < 21.5 \\
    0.50, & \ell_0/\sigma \ge 21.5.  \\
    \end{cases}
\end{equation}

\begin{figure}
	\centering
	\includegraphics[width=.475\textwidth]{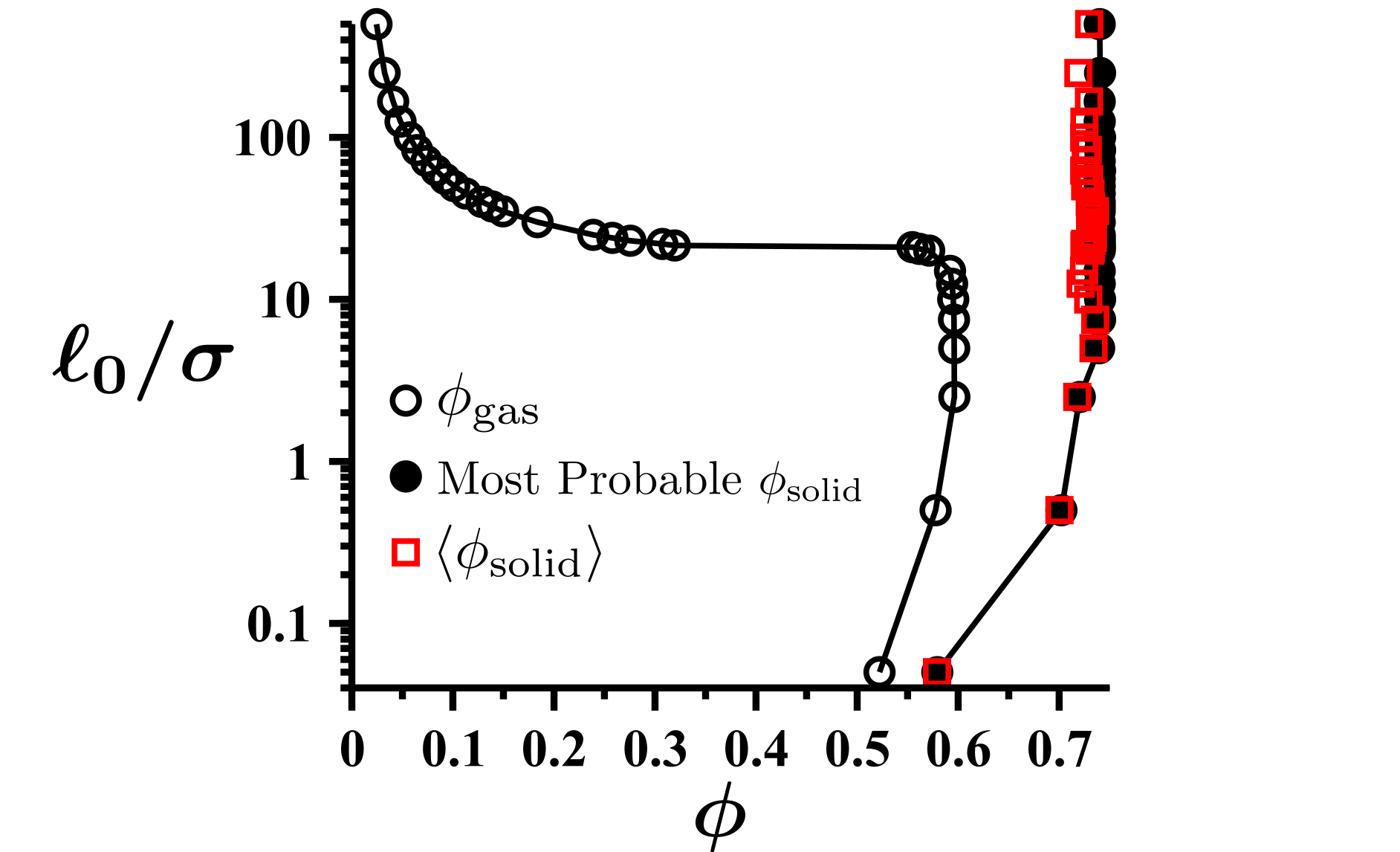}
	\caption{\protect\small{{Comparison of the solid density obtained by using the most probable local volume fraction (presented in the main text) with that obtained from a local volume average $\langle \phi_{\rm solid} \rangle$.}}}
	\label{sm_figure6}
\end{figure}

Obtaining the coexisting densities from eq.~\eqref{eq:dprofile} is problematic as near close-packing crystals (such as those observed for nearly all activities) have a rough density profile associated with the lattice spacing.
We therefore take a different approach and compute distribution of the per-particle volume fraction $P(\phi)$.
As shown in the main text [Figs.~1(c) and 3 in the main text], $P(\phi)$ reveals a sharp peak corresponding to the crystal phase that, for nearly all activities, is the most probable density.
On a logarithmic scale [Fig.~1(c) in the main text], the crystal phase contains additional shallower peaks at slightly lower densities which correspond to particles that neighbor local defects within the crystal. 
This multimodal distribution of local density is unique to the crystal phase, and results in a distinction between the \textit{most probable density} and the \textit{average density}, in contrast to fluids whose distribution are Gaussian.
Figure~\ref{sm_figure6} presents a comparison of the coexisting solid density defined by taking the average of the local crystal density distribution and that defined by defining the solid density to be the most probable density.
The relatively low defect density in these crystals is clear as the inclusion of defects is found to only marginally reduce the crystal density. 
The solid density presented in the main text is defined as the most probable density within the crystal.

\subsection{Triple Point}
The triple point activity was determined by locating the point in which the solid-fluid and liquid-gas phase boundaries intersect. 
As this intersection occurs between activities of $\ell_0/\sigma = 21.0$ and $21.5$ (the fluid density coexisting with the solid phase makes an abrupt shift between these activities) we take the triple point activity to be $\ell_0/\sigma \approx 21.25$.  
The densities of the three coexisting phases [shown in Fig.~1(a) in the main text] are simply the interpolated gas, liquid, and solid densities at $\ell_0/\sigma \approx 21.25$.

\section{Metastable Fluid Details}
Interrogating the nucleation landscape of active fluids near maximal packing requires a careful fluid preparation procedure so as not to tilt the nucleation landscape.
To prepare the fluids presented that were the subject of Fig.~2 in the main text, we first briefly simulate the system at a density of $\phi=0.2$ followed by isotropically compressing the system at a rate of $0.05~U_0$ until the desired $\phi$ is reached. Simulations were performed in a cubic simulation box with $55296$ particles.

\begin{figure}
	\centering
	\includegraphics[width=.475\textwidth]{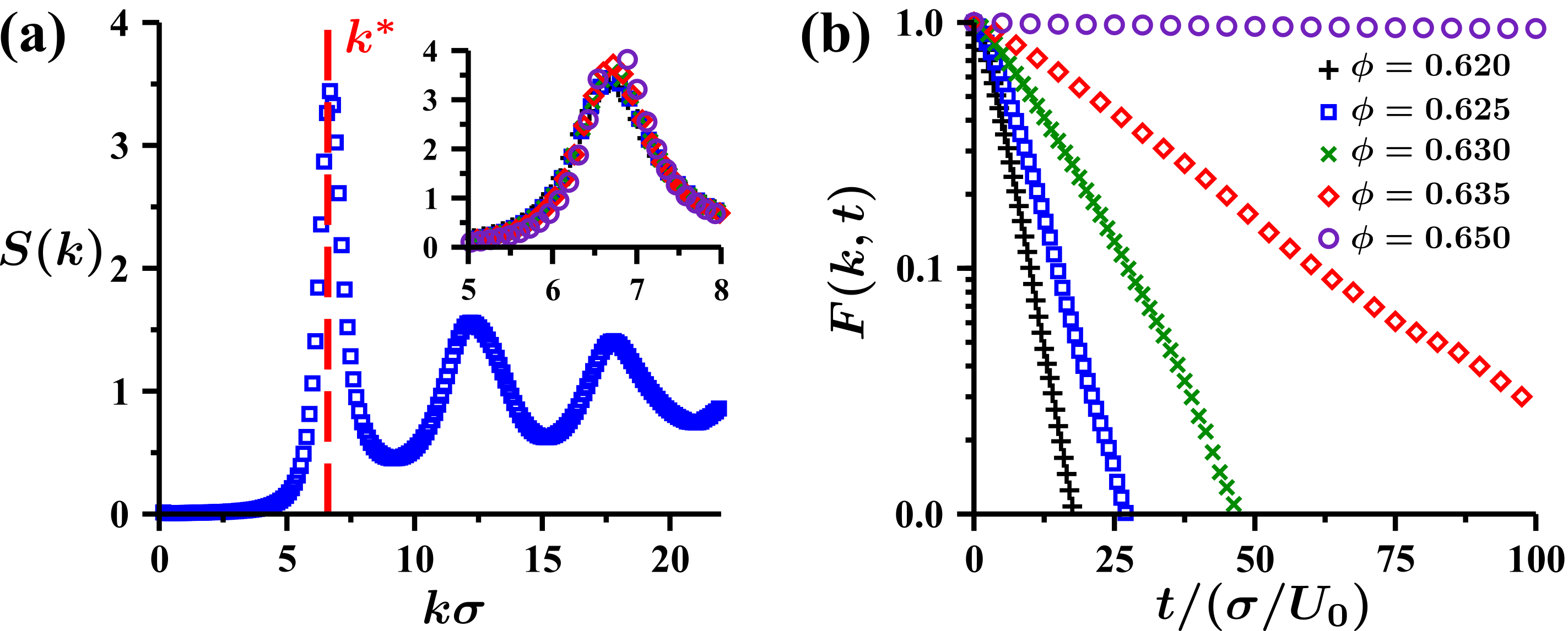}
	\caption{\protect\small{{The (a) static and (b) dynamic structure factors for metastable fluids with activity $\ell_0/\sigma=50$. The inset in (a) highlights the relative concentration [see legend in (b)] independence of the location of the first peak of $S(k)$. }}}
	\label{sm_figure7}
\end{figure}

The fluids prepared by this method showed no evidence of long-range order, made evident with the the static structure factor:
\begin{equation}
\label{eq:Sk}
S(\mathbf{k}) = \frac{1}{N}\biggl\langle\biggl|\sum_{j=1}^N|\exp(\mathrm{i}\mathbf{k}\cdot \mathbf{x}_j)\biggr|^2\biggr\rangle,
\end{equation}
shown in Fig.~\ref{sm_figure7}(a), 
The first peak in the structure factor occurs at a wavevector magnitude of $k^*\sigma \approx 6.3$ a value that coincides the effective hard-sphere diameter (i.e.,~$2^{1/6}\sigma k^* \approx 2\pi$) and is relatively insensitive to the densities examined [see Fig.~\ref{sm_figure7}(a) inset].
Evaluating the self-component of the dynamic structure factor: 
\begin{equation}
\label{eq:selfFk}
F(\mathbf{k}, t) = \frac{1}{N}\biggl\langle \sum_{j=1}^N\exp\left(\mathrm{i} \mathbf{k}\cdot [\mathbf{x}_j(t) - \mathbf{x}_j(0)]\right)\biggr\rangle,
\end{equation}
at $k^*$ we find rapid Arrhenius relaxation for all $\phi < 0.65$, shown clearly in Fig.~\ref{sm_figure7}(b).
The absence of vitrification in these near maximally packed fluids is consistent with recent studies that demonstrate that activity shifts the glass transition to higher $\phi$~\cite{Ni2013, Berthier2014, Flenner2016, Berthier2017, Berthier2019, Nandi2018, Takatori2020}.
Moreover, that nucleation occurs on timescales that are much larger than the characteristic fluid relaxation time (for all $\phi<0.65$) indicates that the fluids are truly metastable (i.e.,~crystallization does not occur before the fluid can relax) providing further confidence in our preparation procedure.